\documentclass[12pt]{article}

\usepackage{latexsym}
\usepackage{amssymb}
\usepackage{graphicx}
\usepackage{amsmath,cite}

\newcommand{\ed}{\end{document}}
\newcommand{\beq}{\begin{equation}}
\newcommand{\eeq}{\end{equation}}

\title{Path Integral Quantization corresponding to the Deformed Heisenberg Algebra}

\begin{document}

\maketitle
\begin{center}
Souvik Pramanik$^\dagger$\footnote{E-mail: souvick.in@gmail.com}, 
Mohamed Moussa$^\ddagger$\footnote{E-mail: mohamed.ibrahim@fsc.bu.edu.eg},
Mir Faizal$^*$\footnote{E-mail: f2mir@uwaterloo.ca} and 
Ahmed Farag Ali$^\ddagger$\footnote{E-mail:ahmed.ali@fsc.bu.edu.eg}

\vspace{0.2 cm}
\small{\emph{$^\dagger$Physics and Applied Mathematics Unit,\\
Indian Statistical Institute, 203 B. T. Road, Kolkata 700108, India}} \\
\small{\emph{$^*$Department of Physics and Astronomy, \\
University of Waterloo, Waterloo, Ontario, N2L 3G1, Canada}}\\
\small{\emph{$^{\ddagger}$Department of Physics, Faculty of Sciences, \\
Benha University, Benha 13518, Egypt}}
\end{center}
\vspace{0.5cm}
	
\begin{abstract}
In this paper, the deformation of the Heisenberg algebra, consistent with both the generalized uncertainty principle and doubly special relativity, has been analyzed. It has been observed that, though this algebra can give rise to fractional derivative terms in the corresponding quantum mechanical Hamiltonian, a formal meaning can be given to them by using the theory of harmonic extensions of function. Depending on this argument, the expression of the propagator of the path integral corresponding to the deformed Heisenberg algebra, has been obtained. In particular, the consistent expression of the one dimensional free particle propagator has been evaluated explicitly. With this propagator in hand, it has been shown that, even in free particle case, normal generalized uncertainty principle and doubly special relativity shows very much different result.
\end{abstract}  


\vskip .3cm

\section{Introduction}

According to the Heisenberg uncertainty principle, it is possible to detect the position of a particle with arbitrary accuracy, if there is no restriction on its momentum. Thus, the fuzziness of the position measurement within a minimum length scale, is not consistent with the usual Heisenberg uncertainty principle. However, almost all approaches to quantum gravity predict the existence of a minimum  measurable length scale. This is because, in most of these approaches the continuum picture of space-time breaks down near the Planck scale, which in turn implies that the Planck length acts like a minimum measurable length scale for all these approaches to quantum gravity. Indeed, there are strong indications from the physics of black holes, that a minimum length of the order of the Planck length should act as an universal feature of all models of quantum gravity \cite{z4,z5}. The reason behind this is the energy needed to probe spacetime below Planck length scale will give rise to a mini black hole in that region of spacetime. Also, in the context of perturbative string theory, the strings are the smallest probe available, and so, it is not possible to probe the spacetime below the string length scale. Thus, the string length scale acts as a minimum measurable length in string theory \cite{z2,zasaqsw,csdcas,cscds,2z}. Even in loop quantum gravity, it is expected that a minimum length scale may exist, and this can have important phenomenological consequence. In fact, in loop quantum gravity, the existence of the minimum length scale turns big bang into a big bounce \cite{z1}.

In order to incorporate the existence of a minimum measurable length scale with the uncertainty relation, the usual Heisenberg uncertainty relation has been generalized to a new uncertainty principle, so called, the Generalized Uncertainty Principle (GUP) \cite{z2,zasaqsw,csdcas,cscds,2z,14,17,18,51,54}. As the usual uncertainty relation is closely related to the canonical Heisenberg algebra, so, this deformed uncertainty principle modifies the canonical Heisenberg algebra to a non-canonical one, so that, the commutator of the position and momentum operators becomes momentum dependent, instead of a constant. With this non-canonical algebra, the coordinate representation of the momentum operator get modified, and this in turn produces correction terms for all quantum mechanical systems. On the other hand, there is a another kind of deformation of the Heisenberg algebra, that has been proposed in the context of Doubly Special Relativity (DSR) \cite{2,21,3} and has the motivation of studying both the velocity of light and Planck energy as universal constants. It may be noted that, the deformed  Heisenberg algebra studied in DSR theory has been predicted from many consequences, such as, discrete spacetime \cite{1q}, spontaneous symmetry breaking of Lorentz invariance in string field theory \cite{2q}, spacetime foam models \cite{3q}, spin-network in loop quantum gravity \cite{4q}, non-commutative geometry \cite{5q}, and Horava-Lifshitz gravity \cite{6q}. It is possible to generalize the DSR in curved space-time and the resultant theory is called the gravity's rainbow \cite{n1,n2}. 

Interestingly, the above two deformation's of the  Heisenberg algebra can be combined into a single deformation of the Heisenberg algebra
\cite{main1,main2}. It can be seen that the commutator of the position and momentum operators in this algebra contains a linear power of momentum, and therefore, called as the linear GUP model. The momentum operators for this model, contain non-local fractional derivative terms in all the quantum mechanical Hamiltonian. However, such non-local terms do not occur for one dimensional systems. This advantage has motivated the study of the deformation of various one dimensional quantum mechanical systems by the linear GUP. The effect of this deformation on the transition rate of ultra cold neutrons in gravitational field has been studied in \cite{n6}. This deformation has been studied in cosmology, turns the big bang into a big bounce  \cite{z6}. Furthermore, the Lamb shift and Landau levels have 
also been analyzed using this deformed Heisenberg algebra \cite{n7}. One of the most interesting result has been obtained from this algebra, is that, the space actually has to be a discrete structure \cite{main1}. In demonstrating the existence of a discrete structure for the space, therein the Schr\"{o}dinger wave equation for a free particle was deformed by linear GUP. Similar results have been obtained by repeating this analysis using the relativistic wave equation by the linear GUP \cite{main2}. Thermodynamical properties of simple quantum mechanical systems have also been analyzed using this deformed Hesinberg algebra \cite{m4}.
It was observed that the corrections to the thermodynamic properties of system caused by this deformed Hesinberg algebra are similar to the corrections produced  from the polymeric quantization. In a series of papers, various applications of the new model of the linear GUP have been investigated 
\cite{applications,bba1, bba122, bba2, bba4,bba5,bba6, bba7, bba8}. A review of the phenomenology  consequences of GUP can be found in \cite{aam, a9, a10}.

Even though a lot of work has been done with both the quadratic GUP and the linear GUP and almost all these results have been derived using canonical quantization. However, the link to classical mechanics in path integral formalism is very intuitive. This is because, in path integral formalism the quantum theory is constructed 
by adding quantum corrections to the classical results. The path integral formalism can also be used for analyzing 
non-perturbative phenomena like instantons in quantum mechanics. The generalization of such non-perturbative phenomena 
to field theory has led to the discovery of some very important results in quantum gravity \cite{a1, a2, a4, a5}.
Furthermore, sometimes in the certain systems like superconductors \cite{super} and super-fluids \cite{super1}, the 
degrees of freedom of these systems lock to form a single collective variables. Path integral is ideally suited to 
deal with such situations \cite{0path, 1path}. So, it is desirable to analyze 
the deformed Heisenberg algebra using the path integral formalism. In fact, recently deformed Heisenberg algebra
corresponding to quadratic  GUP has been analyzed by using the path integral formalism \cite{path}. 
Now, as there are theoretical reasons for the existence of the linear GUP,  
it is important  to extend this work for the  deformation Heisenberg algebra which is consistent linear GUP. This is what we are aimed to do in this paper. First, we will derive the modified propagator of path integral with the linear GUP model, and compare our results with the quadratic GUP and the DSR models. With these comparison in hand, we will show that different deformations give rise to different behavior for this system. In particular, by comparing the graphical plots of probability amplitude of the propagator corresponding to the different algebras, we will explore that they provide different results for different deformation. The probability amplitude decreases for quadratic GUP case, whereas it increases for the linear GUP.  

In section 2,  we will use the harmonic extension of functions to  give  a  formal meaning to the fractional derivative terms that occurs because of the linear GUP.  
In section 3, we will   explicitly derive the expression  for a one dimensional 
free particle propagator using the path integral formalism.   In section 3, we will analyze the properties of  this deformed propagator. Finally, we will summarize our results in section 4. 

\section{Harmonic analysis with the deformed Heisenberg algebra}

In this section, we analyze the non-local Heisenberg algebra using harmonic extension of functions. First, we start with 
the deformed Heisenberg algebra corresponding to the GUP that can be written as \cite{z4,z5,z2,zasaqsw,csdcas,cscds,2z}
\begin{equation}
 [\tilde x^i, \tilde p_j ] =
i \hbar[\delta^i_{j}(1 + \beta \tilde p^2)+ 2 \beta \tilde p^i \tilde p_j],
\end{equation}
where $\beta$ is a parameter of deformation. On the other hand, the deformation of the Heisenberg algebra motivated by DSR can be written as 
\begin{equation}
 [\tilde x^i, \tilde p_j] = i \hbar [\delta^i_{j}(1 - \alpha \sqrt{\tilde p^k \tilde p_k})  + \alpha^2 \tilde p^i \tilde p_j].
\end{equation}
It is possible to unify the above two deformations into a single deformation of the Heisenberg algebra \cite{main2}, given by
\begin{equation}
 [\tilde x^i, \tilde p_j] = i\hbar \left[  \delta^i_{j} - \alpha \sqrt{\tilde p^k \tilde p_k} \delta^i_{j}+ 
 \alpha(\sqrt{\tilde p^k \tilde p_k})^{-1} \tilde p^i \tilde p_j+ \alpha^2 (\tilde p^k\tilde p_k) \delta^i_{j}
  + 3\alpha^2 \tilde p^i \tilde p_j\right],\label{x-p-com}
\end{equation}
where $\alpha$ is the deformation parameter defined by $\alpha = {\alpha_0}/{M_{Pl}c} = {\alpha_0 \ell_{Pl}}/{\hbar}$, 
$M_{Pl}$ is the Planck mass, $\ell_{Pl}$ ($\approx 10^{-35}~m)$ is the Planck length, $M_{Pl} c^2$ ($\approx 10^{19}~GeV)$ 
is the Planck energy. The upper bounds on the parameter $\alpha_0$ has been calculated in \cite{n7}, and it was proposed that this
GUP may introduce an intermediate length scale between Planck scale and electroweak scale. Recent proposals suggest that
this bound can be measured by using quantum optics techniques \cite{Nature}, and also by gravitational wave techniques \cite{NatureGRW},
which are to be considered as a milestone in quantum gravity phenomenology. 

It may be noted that in one dimension, the deformed uncertainty relation corresponding to this algebra,  can be written as 
$\Delta \tilde x \Delta \tilde p \geqslant \frac{\hbar}{2}[1 - 2 \alpha \langle \tilde p \rangle + 4 \alpha^2 
\langle \tilde p^2\rangle ]$. This in turn implies the existence of a minimum length $\Delta \tilde x_{min} \approx 
\alpha_0 \ell_{Pl}$ and a maximum momentum  $ \Delta \tilde p_{max}\approx \alpha_0^{-1} M_{Pl} c$. Now, the deformed commutation relation (\ref{x-p-com}) modifies the momentum operator as $\tilde p_i = p_i (1 - \alpha  \sqrt{ p^k p_k}
+ 2\alpha^2  p^k    p_k )$, where
$ \tilde x_i =    x_i $, and the canonical pair $(x,p)$ satisfies the canonical algebra  $[   x^j,    p_i] = i \hbar \delta^j_i$. Here $   p_i$ can be interpreted as the momentum
at low energy. This deformed momentum operator modifies the original Hamiltonian
\begin{equation}
	H= \frac{1}{2 m}    \tilde p^i \tilde p_i + V (\tilde x),
\end{equation}
to
$H = H_o + H_d,$ where the correction Hamiltonian $H_d$ is given by
\begin{equation}
	H_d = -\frac{\alpha}{m}    p^i    p_i \sqrt{   p^j    p_j} + \frac{5 \alpha^2}{2m}    (p^i    p_i)    (p^j    p_j).
	\label{ham-gen}
\end{equation}
Since, the standard coordinate representation of the canonical momentum $p$ is given by $ p_i = -i\hbar\partial_i$, so, the coordinate
representation for $ \tilde p_i$ can be written as
\begin{equation}
\tilde p_i = -i\hbar \left(1 - \alpha\hbar \sqrt{-\partial^j \partial_j} - 2\alpha^2 \hbar^2\partial^j
\partial_j\right)\partial_i.
\end{equation}
Using this form of momentum operator we obtain the deformed Schr\"{o}dinger equation as
\begin{equation}
- \frac{\hbar^2}{2m} \partial^i\partial_i  \psi + \frac{\alpha\hbar^3}{m}  \partial^i\partial_i \sqrt{-\partial^j\partial_j}\psi+ 
\frac{5 \alpha^2\hbar^4}{2m}
\partial^i\partial_i \partial^j\partial_j \psi +  V(x)\psi =i\hbar\partial_t  \psi. \label{sch}
\end{equation}

Notice that, there are  fractional derivative terms present in the deformed Schr\"{o}dinger equation (\ref{sch}). A formal meaning can be given to these terms in the framework of harmonic extensions of a function   \cite{lifz2,24,sdcvadf,ahusxua, lifz}. We start with by defining a harmonic function $u: R^n \times (0, \infty) \to R$,  such that, its restriction  to $R^n$  coincides with  
the wave  function of the deformed Schr\"{o}dinger equation, $ \psi   : R^n \to R$. It is possible to find $u$ by solving 
the Dirichlet problem defined by: $ \partial^2_{n+1} u (x, y) =0  $ with $u(x, 0) = \psi (x) $, for a given wave function 
$ \psi $. Here $\partial^2_{n+1}$ is defined to be the Laplacian in $R^{n+1}$, such that  $x \in R^n$ and $ y \in R$. So, 
there is a unique harmonic extension  $ u \in C^\infty (R^n \times (0, \infty))$ for a smooth function $C^\infty_0 (R^n) $.
We can now give  a formal meaning to  $\sqrt{-\partial^j\partial_j}$ by analyzing its action on the wave function $ \psi   
: R^n \to R $, such that the harmonic 
extension of the wave function $u: R^n \times (0, \infty) \to R$ satisfies,
 \begin{equation}
\sqrt{-\partial^j\partial_j}    \psi (x)  = - \frac{\partial u (x, y)}{ \partial y}\left. \right| _{y =0}.
\end{equation}
It may be noted that $ u_y (x, y)$ is the harmonic extension of $\sqrt{-\partial^j\partial_j}\psi (x)$ to $R^n \times (0,
\infty)$. This is because $u: R^n \times (0, \infty) \to R$ is a harmonic extension of the wave function  $ \psi  
: R^n \to R $. The successive application  of $\sqrt{-\partial^j\partial_j}$ on the wave function is given by
\begin{eqnarray}
\left[\sqrt{-\partial^j\partial_j}\right]^2\psi  (x) &=& \frac{\partial^2 u(x, y)}{\partial y^2}\left. \right|_{y =0}
\nonumber \\
&=& - \partial ^2_n u(x, y)\left. \right|_{y =0}.
\end{eqnarray}
This equation can be used to give a formal meaning to  $ \sqrt{- (\partial^k \partial_k )}$. Now  as $u (x, y)$ is the
harmonic extension of the wave function $\psi (x)$, and $\sqrt{-\partial^j\partial_j} \psi = - u_y (x, 0) $,
therefore the harmonic extension of  $\partial_i    \psi (x) $  can be written as $ \partial_i u (x, y)$. Now if 
$u \in C^2 (R \times (0, \infty))$, we also have
\begin{equation}
\sqrt{-\partial^j\partial_j} \partial_i    \psi (x)  = - \partial_i u_y (x, y) \left. \right|_{y =0}.
\end{equation}
Thus, this non-local operator $\sqrt{-\partial^j\partial_j}$   commutes with a derivative,
 \begin{equation}
  \sqrt{-\partial^j\partial_j} \partial_i    \psi (x)  = \partial_i \sqrt{-\partial^j\partial_j}    \psi (x) .
 \end{equation}
For example, we can analyze the action of  $\sqrt{-\partial^j\partial_j} $ on $\exp(ikx)$, using the framework of
harmonic extensions of a function.  First, let  we consider $   \psi (x)  = \cos kx$. Its   harmonic extension
can be written as  $u (x, y) = \exp-|k| y .\cos kx$, where $ y \in (0, \infty)$. Therefore, we can write
\begin{equation}
\frac{\partial^2 u (x, y) }{\partial^2 x } + \frac{\partial^2 u(x, y)}{\partial^2 y } = 0
\end{equation}
Thus, the action of $\sqrt{-\partial^j\partial_j} $ on the wave function $ \psi$ can be evaluated as
\begin{equation}
\sqrt{-\partial^j\partial_j}  \cos kx = - u_y (x, y) \left. \right|_{y =0}.
\end{equation}
So, we obtain the following result $\sqrt{-\partial^j\partial_j}  \cos kx =|k| \cos kx$. A similar result can be 
obtained by considering,    $ \psi    = \sin (kx)$. This is because its   harmonic extension on $R^n \times (0, \infty)$
can be written as
$u (x, y) =  \exp -|k| y . \sin kx$, and so, we obtain $\sqrt{-\partial^j\partial_j}  \sin kx =|k| \sin (kx)$. Finally,
the action of  $\sqrt{-\partial^j\partial_j} $  on $\exp ikx$ can be written as
\begin{equation}
\sqrt{-\partial^j\partial_j} \exp ikx = \sqrt{-\partial^j\partial_j} (\cos kx + i \sin kx).
\end{equation}
Thus, we obtain the desired result,  $\sqrt{-\partial^j\partial_j} \exp ikx = |k| \exp ikx$.

Now we define $ u_1 (x, y)$ as the  harmonic extensions of $\psi_1 (x)$ to $ C = R^n \times (0, \infty)$, and
$u_2(x, y)$ as the harmonic extensions of $\psi_2(x)$ to $ C = R^n \times (0, \infty)$.
We also assume that both these  harmonic extensions vanish for $|x| \to \infty $ and $|y| \to \infty $. It is now possible 
to write \cite{24},
\begin{equation}
\int_C u_1(x, y) \partial_{n+1}^2 u_2 (x, y) dx dy -
\int_C  u_2(x, y) \partial_{n+1}^2 u_1 (x, y) dx dy
= 0.
\end{equation}
So, we obtain the following expression
 \begin{equation}
 \int_{R^n} \left(u_1(x, y) \partial_y u_2 (x, y)  -   u_2(x, y)
\partial_x u_1 (x, y) \right)\left. \right|_{y =0} dx
= 0.
 \end{equation}
This expression can be written in terms of the functions $ \psi_1 (x) $ and $ \psi   _2 (x)$ as,
\begin{equation}
 \int_{R^n}\left( \psi   _1(x) \partial_y
\psi _2 (x) - \psi _2(x)\partial_x\psi _1 (x)\right)  dx
= 0.
 \end{equation}
Thus, the non-local differential operator  $\sqrt{-\partial^j\partial_j}$ can be moved from $ \psi   _2 (x)$ to $ \psi   _1 (x)$,
 \begin{equation}
 \int_{R^n}\psi _1 (x) \sqrt{-\partial^j\partial_j}\psi _2 (x)~dx =
 \int_{R^n}\psi _2 (x) \sqrt{-\partial^j\partial_j}\psi _1 (x)~dx.
 \end{equation}

So far, we have analyzed a general deformed Heisenberg algebra consistent with both GUP and DSR, and demonstrated that, even though
this algebra could give rise to fractional derivative terms in the Hamiltonian, but it is possible to formally deal with these
non-local terms, in light of the harmonic extension of functions. We now derive an expression for the quantum mechanical propagator 
corresponding to this deformed Heisenberg algebra. First, note that  in the deformed Schr\"{o}dinger equation, the modified 
Hamiltonian does not have any explicit time dependence,
\begin{equation}
 H  \psi = i\hbar \frac{\partial }{\partial t } \psi,
\end{equation}
where
\begin{equation}
 H  = \frac{1}{2 m}    p^i    p_i  -\frac{\alpha}{m}    p^i    p_i \sqrt{   p^j    p_j} + \frac{5 \alpha^2}{2m}   p^i    p_i \ p^j    p_j + V (x) \label{ham-3} .
\end{equation}
So, if we know the wave function $ \psi (x, t')$, then we can explicitly write the wave function $\psi (x,t'')$, using the 
propagation relation
\begin{equation}
 \psi (x, t'') = \exp  [-i H( t'' - t'  )/\hbar] \ \psi (x,t').
\end{equation}
For a small time interval $\Delta t=t''-t'$, we have
\begin{eqnarray}
 \langle x | \exp (- i H \Delta t/\hbar)  | x \rangle &=&
 \int d p  \langle x| e^{\left[- \frac{i}{\hbar}\left( \frac{1}{2m}    p^i    p_i  -\frac{\alpha}{m}
 p^i    p_i \sqrt{   p^j    p_j} + \frac{5 \alpha^2}{2m}   p^i    p_i    p^j    p_j\right) \Delta t \right] }|p   \rangle
  \langle p | e^{-\frac{ i}{\hbar}  V(x) \Delta t}  |x  \rangle \nonumber \\ &=&
 \int \frac{dp}{2\pi\hbar} e^{\left[- \frac{i}{\hbar} \left( \frac{1}{2 m}    p^i    p_i  -\frac{\alpha}{m}
 p^i    p_i \sqrt{   p^j    p_j}+ \frac{5 \alpha^2}{2m}   p^i    p_i    p^j    p_j + V (x) \right)\right]  \Delta t}. \label{pro-0}
\end{eqnarray}
Therefore, the quantum mechanical propagator for small time interval $\Delta t=t''-t'$ corresponding to this non-local
Hamiltonian can be written as
\begin{equation}
K(x'',t'';~x',t')=\int e^{\frac{i}{\hbar}\int_{t'}^{t'+\Delta t} \mathcal{L}   d t}~
	\frac{d   {p}}{2\pi\hbar},  \label{ker-free}
\end{equation}
where $ \mathcal{L} =  {p} ^i.\dot{x}_i
	-\frac{  {p}^ i p_i }{2m}+\frac{\alpha}{m}   {p}^i p_i \sqrt{p^j p_j }-\frac{5 \alpha^2}{2 m}  {p}^ i p_i~p^j p_j 
	- V(x) .$

\section{ Explicit form of propagator }

In the previous section, we have given formal description of the propagator corresponding to the deformed Heisenberg algebra 
consistent with both GUP and DSR. However, it is computationally very difficult to calculate the full
three dimensional form of this propagator. So, in this section, we explicitly calculate the one dimensional free particle
propagator corresponding to this deformed Hamiltonian. The above form of propagator (\ref{pro-0}) can be expanded up to the 
second order of $\alpha$ as,
\begin{equation}
K(x'',t'';x',t') = \int\frac{dp}
{2\pi\hbar}  e^{-\frac{i \Delta t}{2 m \hbar}\left(p-\frac{(x''-x')m}{\Delta t}\right)^{2}+\frac{i m(x''-x')^{2}}{2\hbar\Delta t}} 
\times
\left(1+\frac{i\alpha \Delta t}{m \hbar}p^3-\frac{5 i \alpha^2 \Delta t}{2 m \hbar}p^4-\frac{\alpha^2 \Delta t^2}{2 m^2 
\hbar^2}p^6\right). \label{ker-10}
\end{equation}
Integrating this, we get the expression for the propagator for infinitesimal time interval $\Delta t=t''-t'$ as,
\begin{eqnarray}
K(x'',t''~;~x',t') &=& \sqrt{\frac{m}{2\pi i\hbar\Delta t}}
\left[1+\alpha \left(\frac{3 m (x''-x')}{\Delta t}+\frac{i m^2 (x''-x')^3}
{\hbar \Delta t^2}\right)\right.\nonumber\\ &&\left.
+\alpha^2 \left(
\frac{15 m^2 (x''-x')^2}
{2 \Delta t^2}+\frac{5 i m^3 (x''-x')^4}
{\hbar \Delta t^3}- \frac{m^4 (x''-x')^6}{2 \hbar^2 \Delta t^4}\right)\right]
 \times e^{\dfrac{im(x''-x')^{2}}{2\hbar\Delta t}}.  
 \label{ker-12}
\end{eqnarray}

Note that, if we consider the terms up to third order of $O(\alpha)$, then the one dimensional  deformed  Hamiltonian can 
be written as 
\begin{equation}
 H  = \frac{1}{2m}p^2 -\frac{\alpha}{m} p^3  + 
\frac{5 \alpha^2}{2m} p^4 +\frac{2\alpha^3}{m}p^5 +V (x).
\end{equation}
Using this deformed Hamiltonian, the propagator to the third order of $O(\alpha)$, can be written as 
\begin{eqnarray}
K(x'',t''~;~x',t') &=& \sqrt{\frac{m}{2\pi i\hbar\Delta t}}
\left[1+\alpha \left(\frac{3 m (x''-x')}{\Delta t}+\frac{i m^2 (x''-x')^3}
{\hbar \Delta t^2}\right) \right.\nonumber\\&&\left.  +\alpha^2 \left(
\frac{15 m^2 (x''-x')^2}
{2 \Delta t^2}
+\frac{5 i m^3 (x''-x')^4}
{\hbar \Delta t^3}- \frac{m^4 (x''-x')^6}{2 \hbar^2 \Delta t^4}\right) \right.\nonumber\\&&\left. 
+\alpha^3 \left(-\frac{135 i \hbar m^2 (x''-x')}
{\Delta t^2}
-\frac{145 m^3 (x''-x')^3}
{2 \Delta t^3}\right.\right.\nonumber\\&&\left.\left. +\frac{25 i m^4 (x''-x')^5}{2 \hbar \Delta t^4}-\frac{7 m^5 (x''-x')^7}
{2 \hbar^2 \Delta t^5}-\frac{i m^6 (x''-x')^9}
{6\hbar^3 \Delta t^6}\right)\right] \times
e^{\dfrac{im(x''-x')^{2}}{2\hbar\Delta t}}. \label{ker-13}
\end{eqnarray}
Proceeding in this way one can construct more general form of the combined GUP and DSR corrected free particle propagator.
 
Note that, the path integral corresponding to polynomial term $p^4$, that comes from ordinary GUP contribution, has already been analyzed \cite{path}. Therefore in this paper, we are mainly interested in analyzing the effect of the linear corrections to the Heisenberg algebra, that comes from the DSR. From the Hamiltonian (\ref{ham-3}), one can observe that the terms containing linear order correction of $\alpha$ come from DSR theory. So, we neglect terms proportional to $\alpha^2$ 
for explicit calculations. Thus, here we concentrate on the part of  the propagator, containing first order of $\alpha$, i.e.
\begin{eqnarray}
K(x'',t''~;~x',t') &=& \sqrt{\frac{m}{2\pi i\hbar\Delta t}}
\left[1+\frac{3 \alpha m (x''-x')}{\Delta t}+\frac{i \alpha m^2 (x''-x')^3}
{\hbar \Delta t^2}\right]e^{\frac{im(x''-x')^{2}}{2\hbar\Delta t}}. \label{ker-5}
\end{eqnarray}
In order to obtain free particle propagator for a finite time interval $t''-t'$, we divide the interval into $N$ subintervals 
of equal length $\Delta t$ such that $t''-t'=N \Delta t$ and use the above result given by Eq. (\ref{ker-5}) for each subinterval. 
Therefore the propagator for finite time interval can be written as
\begin{eqnarray}
K(x'', t''; x', t') &=& \left(\sqrt{\frac{m}{2 \pi i \hbar \Delta t}}\right)^N \int d x_1 d x_2 ...d x_{N-1}~e^{\frac{i m}{2 \hbar
\Delta t}
[(x_1 - x_0)^2 + (x_2-x_1)^2+...+(x_N - x_{N-1})^2]} \nonumber\\
&&\times
\left\{1+\frac{3 \alpha m (x_1-x_0)}{\Delta t}+\frac{i\alpha m^2
(x_1-x_0)^3}{\hbar \Delta t^2}\right\} \times
\left\{1+\frac{3 \alpha m (x_2-x_1)}{\Delta t}+\frac{i\alpha m^2 (x_2-x_1)^3}{\hbar \Delta t^2}\right\}\nonumber\\
&&\times ...\times \left\{1+\frac{3 \alpha m (x_N-x_{N-1})}{\Delta t}
+\frac{i\alpha m^2(x_N-x_{N-1})^3}{\hbar \Delta t^2} \right\}. \label{ker-6}
\end{eqnarray}
Neglecting the term containing higher order of $\alpha$, we obtain the reduced expression
\begin{eqnarray}
K(x'',t'';x',t') &=& \left(\sqrt{\frac{m}{2\pi i\hbar \Delta t}}\right)^N\int d x_1...d x_{N-1}\times e^{\frac{i m}{2 \hbar 
\Delta t}[(x_1-x_0)^2 + (x_2-x_1)^2+    ...+(x_N - x_{N-1})^2]}\nonumber\\
&&\times \left[1+\frac{3 \alpha m}{\Delta t}(x_N-x_0)
+\frac{i\alpha m^2}{\hbar \Delta t^2}\{(x_1-x_0)^3+...+(x_N-x_{N-1})^3\}\right].
\label{ker-7}
\end{eqnarray}
The integrations present in this expression can be calculated as,  
\begin{equation}
\int~ e^{~i\lambda \left[(x_1-x_0)^2+~...~+(x_N-x_{N-1})^2\right]} ~dx_1~dx_2~...~dx_{N-1}
=\frac{1}{\sqrt{N}}\left(\frac{i\pi}{\lambda}\right)^{\frac{N-1}{2}}~e^{\frac{i  \lambda(x_N-x_0)^2}{N}},\label{int-1} 
\end{equation}
and 
\begin{eqnarray}
&&\int~ e^{~i\lambda \left[(x_1-x_0)^2+~...~+(x_N-x_{N-1})^2\right]} \left[(x_1-x_0)^3+~...
~+(x_N-x_{N-1})^3\right]~dx_1~dx_2~...~dx_{N-1}\nonumber\\
&& = \frac{\lambda}{i N^2\sqrt{N}}
\left( \dfrac{i\pi}{\lambda} \right)^{\frac{N-1}{2}}(x_N-x_0)\left[-\dfrac{3}{2}N(N-1)+i\lambda (x_N-x_0)^2
\right]~e^{\frac{i  \lambda(x_N-x_0)^2}{N}}.\label{int-2}  
\end{eqnarray}
Substituting (\ref{int-1},\ref{int-2}) into (\ref{ker-7}), the propagator becomes 
\begin{eqnarray}
&& K(x''~,~t'';x'~,~t') \nonumber\\
&& = \sqrt{\dfrac{m}{2\pi i \hbar N\Delta t}} 
\left[ \left(1+3\alpha m\dfrac{x_N-x_0}{\Delta t}\right) + \left( 3\alpha m\dfrac{(1-N)(x_N-x_0)}{N\Delta t} 
+ i\alpha m^2\dfrac{ (x_N-x_0)^3}{ \hbar N^2\Delta t^2} \right) \right]e^{\dfrac{im(x_N-x_0)^2}{2\hbar N\Delta t}} \nonumber\\
&& = \sqrt{\dfrac{m}{2\pi i \hbar N\Delta t}}\left[1+3\alpha m\dfrac{x_N-x_0}{N\Delta t}
+i\alpha m^2\dfrac{ (x_N-x_0)^3}{ \hbar N^2\Delta t^2}\right]e^{\dfrac{im(x_N-x_0)^2}{2\hbar N\Delta t}}. 
\end{eqnarray}
Replacing $x_N$ and $x_0$ by $x''$ and $x'$ respectively, and using $N\Delta t =t''-t'$ we obtain the final expression as  
\begin{equation}
K(x'', t''; x', t') = \sqrt{\frac{m}{2 \pi i \hbar (t''-t')}}
\left[1+\frac{3 \alpha m (x''-x')}{(t''-t')}+\frac{i\alpha m^2(x''-x')^3}{\hbar (t''-t')^2}\right]
e^{\dfrac{i m (x''-x')^2}{2 \hbar (t''-t')}}.
\label{ker-final}
\end{equation}
It may be noted that the final form of propagator given by Eq. (\ref{ker-final}) has exactly the same
form as the propagator for the infinitesimal interval given by Eq. (\ref{ker-5}).\\

Now, if we calculate the probability of detecting the particle at a finite region $\Delta x$, enclosing final point $x''$, from (\ref{ker-final}) we get
\begin{equation}
P(x'')=K^{\ast}(x''~,~t''~;~x'~,~t')K (x''~,~t''~;~x'~,~t')=\left[1+6\alpha m\dfrac{x''-x'}{t''-t'}\right]
\dfrac{m}{2\pi \hbar (t''-t')}.
\end{equation}
It is well known that this probability for the ordinary case is given by $P(x'')=\frac{m}{2\pi \hbar (t''-t')}$.
Thus, the probability amplitude increases with the contribution of DSR in the path integral. This probability obtains a maximum value for the maximum momentum,
\begin{equation}
 m\dfrac{x''-x'}{t''-t'}=p\approx(\Delta{p})_{max}=\dfrac{1}{\alpha}.
\end{equation}
Furthermore, as the measurement is made at scales greater than the Planck scale, the corresponding  momentum 
also decreases away from this maximum value, and this in turn makes the probability to decreased and tend 
back to its original value. Now using the structure of propagator we can calculate the expression for
$|K(x'',t'';x',t')|^2$. If $K(x'',t'';x',t')_{GUP}$ is the propagator obtained by considering only the effects 
of GUP, $K(x'',t'';x',t')_{DSR}$ is the propagator obtained from considering only the effects of DSR, 
and $K(x'',t'';x',t')_{GUP+DSR}$ is the propagator obtained from considering the effects of both GUP and DSR, then we obtain
\begin{eqnarray}
|K(x'',t'';x',t')_{GUP}|^2 dx &=& \frac{m}{2\pi\hbar (t''-t')}(1-12\alpha^2 p^2)~dx, \label{K*K-beta}\\
|K(x'',t'';x',t')_{DSR}|^2 dx &=& \frac{m}{2\pi\hbar (t''-t')}(1+6\alpha p)~dx,\\
|K(x'',t'';x',t')_{GUP+DSR}|^2 dx &=& \frac{m}{2\pi\hbar (t''-t')}(1+6\alpha p+24\alpha^2 p^2)~dx,\\
|K(x'',t'';x',t')_{GUP+DSR}|_{upto~O(\alpha^3)} dx &=& \frac{m}{2\pi\hbar (t''-t')}(1+6\alpha p+24\alpha^2 
p^2-100\alpha^3p^3)~dx,\label{K*K-alpha}
\end{eqnarray}
where, the momentum for a particle has been expressed as  $ p =  {m(x''-x')}/ ({t''-t'})$. In the last
expression, we have used the $O(\alpha^3)$ approximation for combined GUP and DSR deformed propagator. It will 
be very interesting to plot the above expressions graphically. Since, we have taken $\alpha$ as a small parameter, first we can choose it within the interval $0<\alpha\leqslant1$. This inequality gives the bound for $\alpha_0$ as, 
$0<\alpha_0\leqslant6.52 \ ( \equiv M_{pl}c)$. However, since the minimum length is of the order of $\Delta x_{min}\approx\alpha_0 l_{pl}$, 
so, $\alpha_0$ cannot be taken less than $1$. Therefore, the allowed region for $\alpha_0$ becomes, $1\leqslant\alpha_0\leqslant6.52$. This inequality implies back the allowed
region for $\alpha$ as $ {1}/{M_{pl}c} \equiv 0.153 \leqslant\alpha\leqslant1$. Now, as $|K(x'',t'';x',t')|^2\geqslant0$, then from the above expressions (\ref{K*K-beta} - \ref{K*K-alpha}) the particle momentum $p$ can be found within $0\leqslant p\leqslant {0.28}/{\alpha}$. 
The corresponding plot is as follows:

\begin{figure}[htb]
   {\centerline{\includegraphics[width=9cm, height=4cm] {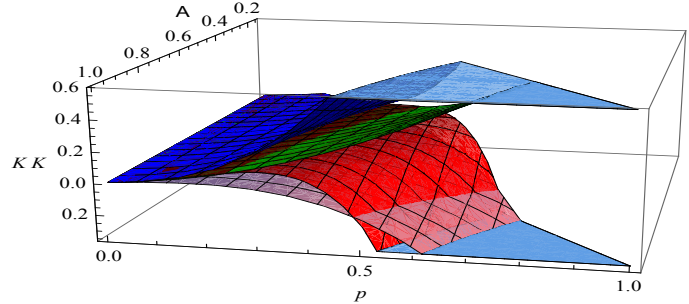}}}
    \caption{{\it{Here the pink plot (lowermost one), the green plot (intermediate), the bule plot (uppermost one) and the red plot (intermmediate) are corresponding to 
    normal GUP case, normal DSR case, combined GUP and DSR case, combined GUP and DSR case upto third order of $\alpha$, respectively.}}} \label{fig1}
 \end{figure}
One can see from the above plot, that for vary high energy momentum $( \approx 0.28/\alpha)$, the difference between the probability amplitudes
of normal GUP and normal DSR case, increases as $\alpha$ takes more and more higher values. But, whenever both of them are combined, the combined probability amplitude is found to be similar to the DSR case. A probable reason behind this is that, the DSR deforms the system to the first order in $\alpha$, and the quadratic GUP deforms the system to the second order in $\alpha$, respectively. This can be seen from the above Eqs. (\ref{K*K-beta}-\ref{K*K-alpha}).
Therefore, it can be conclude that for very higher energies, quadratic GUP behaves differently from the DSR. This is one of our main result in this paper.\\

Now, we show that the propagator corresponding to the deformed Heisenberg algebra satisfies the basic properties of a propagator. Suitably differentiating the above propagator (\ref{ker-final}), with respect to $x''$ and to $t''$, it can be demonstrated that it satisfy the modified Schr\"{o}dinger equation at the final state $(x'',t'')$
\begin{equation}
-\frac{\hbar^{2}}{2m}\frac{\partial^2 K(x'', t''; x', t')}{\partial {x''}^{2}}
+\frac{i \alpha \hbar^3}{m}\frac{\partial^3 K(x'', t''; x', t')}{\partial {x''}^3}=
i\hbar\frac{\partial K(x'', t''; x', t')}{\partial t''}.
\label{ker-con-1}
\end{equation}

Up to the first order of $\alpha$, the deformed Schr\"{o}dinger equation (\ref{sch}) can be written as
\begin{equation}
-\frac{\hbar^{2}}{2m}\frac{\partial^2 \psi(x,t)}{\partial {x}^2}+\frac{i \alpha \hbar^3}{m}\frac{\partial^3 
\psi(x,t)}{\partial x^3}
=i\hbar\frac{\partial\psi(x,t)}{\partial t}. \label{gup-sch}
\end{equation}
The solution of the above Schr\"{o}dinger equation given by Eq. (\ref{gup-sch}) is given by,
\begin{equation}
\psi(x, t) = \left(A e^{i k(1+ \alpha \hbar k)x-\frac{i E t}{\hbar}} +
B e^{-i k(1- \alpha \hbar k)x-\frac{i E t}{\hbar}}
+ C e^{\frac{i x}{2\alpha\hbar}-\frac{i E t}{\hbar}}\right).\label{sol-sch}
\end{equation}
Using this solution Eq. (\ref{sol-sch}), it is possible to show that the
propagator for a small time interval, given by Eq. (\ref{ker-5}), indeed propagates the wave function $\psi(x, t)$ from a initial state $(x', t')$ to the slightly latter state $(x'', t'')$. This means the following relation is 
satisfied, to the first order of $O(\alpha)$:
\begin{equation}
\psi(x'', t'') = \int K (x'', t''; x',t') \psi(x', t') d x'.
\label{pro-1}
\end{equation}
In order to prove the propagation for any finite time interval, we divide the time interval $(t''-t')$ into $N$ 
subintervals of equal length
$\Delta t$ and then apply Eq. (\ref{pro-1}) for each subinterval as,
\begin{eqnarray}
&&\int K(x'', t'';x',t')~\psi(x',t')~dx' \nonumber\\
&=& \int\int..\int K(x'',t'';x_{N-1},t_{N-1})K(x_{N-1},t_{N-1};x_{N-2},t_{N-2})
..K(x_{1},t_{1};x',t')\times \psi(x',t')dx'dx_1...dx_{N-1} \nonumber \\ 
&=& \int\int..\int K(x'',t'';x_{N-1},t_{N-1})K(x_{N-1},t_{N-1};x_{N-2},t_{N-2})..K(x_{2},t_{2};x_{1},t_{1})\times 
\psi(x_{1},t_{1})
dx_{1}...dx_{N-1}
\nonumber\\
&=&.....=\int K(x'',t'';x_{N-1},t_{N-1})\psi(x_{N-1},t_{N-1})dx_{N-1}\nonumber \\
&=& \psi(x'',t''). \label{propfinite}
\end{eqnarray}

Furthermore, one can see that the propagator (\ref{ker-final})
satisfies the normalization criteria of a consistent propagator. This can be shown as follows: taking the complex conjugate
of the Eq. (\ref{ker-final}), we obtain
\begin{equation}
K^{*}(x'', t''; x', t')= \sqrt{\frac{-m}{2 \pi i \hbar (t''-t')}} \left[1+\frac{3\alpha m (x''-x')}{(t''-t')}
-\frac{i\alpha m^2(x''-x')^3}{\hbar(t''-t')^2}\right]
\times e^{\dfrac{-i m (x''-x')^2}{2 \hbar (t''-t')}}.\label{pro*}
\end{equation}
Now, using Eq. (\ref{pro*}),  we can demonstrate, 
\begin{eqnarray}
&& \int K^{\ast}(x'',t''~;~x'_1,t')K(x'',t''~;~x',t')~dx''\nonumber\\
&& =\dfrac{m}{2\pi \hbar (t''-t')}e^{\frac{i m (x'^2-x_1'^2)}{2\hbar(t''-t')}}\int
\left[1-\dfrac{3\alpha m}{(t''-t')}(x'+x_1')+
\left(\dfrac{6\alpha m}{(t''-t')}+\dfrac{3i\alpha m^2 (x'^2-x_1'^2)}{\hbar (t''-t')^2}\right)x''
\right. \nonumber\\
&& \left. ~~-\dfrac{i\alpha m^2}{\hbar (t''-t')^2}(x'^3-x_1'^3)-\dfrac{3i\alpha m^2}{\hbar (t''-t')^2}(x'-x_1')x''^2\right]
e^{\frac{im(x'-x_1') x''}{\hbar (t''-t')}}dx''. 
\end{eqnarray}
If we substitute $u= {m}{}x''/  \hbar (t''-t') $ and use the identity, 
\begin{equation}
\int e~^{-u.(x'-x_1')}u^n~du=2\pi (i)^n \dfrac{d^n \delta(x'-x_1')}{d(x'-x_1')^n}, 
\end{equation}
we obtain the following expression, 
\begin{eqnarray}
&&\int K^{\ast}(x'',t''~;~x'_1,t')K(x'',t''~;~x',t')~dx''\nonumber\\
&& = \left[\left(1-\dfrac{3\alpha m (x'+x_1')}{(t''-t')}-\dfrac{i\alpha m^2 (x'^3-x_1'^3)}{\hbar (t''-t')^2}\right)
\delta(x'-x_1') + \left(6i\alpha \hbar -\dfrac{3\alpha m (x'^2-x_1'^2)}{(t''-t')}\right)\delta'(x'-x_1')\right.\nonumber\\
&& \left. ~~+3i\alpha \hbar(x'-x_1')\delta''(x'-x_1')\right]e^{\frac{i m (x'^2-x_1'^2)}{2\hbar (t''-t')}}.
\label{a6aaaa}
\end{eqnarray}
Multiplying both sides of Eq. (\ref{a6aaaa}) by $\psi^{*}(x_{1}',t')$, then integrating out $x'_1$ by using the identity, 
\begin{equation}
\int \delta^{(n)}(y-x)f(y)dy=-\int \dfrac{\partial f}{\partial x}~ \delta^{(n-1)}(y-x) dy,
\end{equation}
we obtain the following result, 
\begin{eqnarray}
&&\int \int K^{\ast}(x'',t''~;~x'_1,t')K(x'',t''~;~x',t')\psi^{\ast}(x_1',t')dx_1'~dx''\nonumber \\
 && = \left[\left(1-\dfrac{6\alpha m x'}{t}\right)+
\left(\dfrac{12\alpha m x'}{t}+6i\alpha \hbar\right) -\left(\dfrac{6\alpha m x'}{t}
+6i\alpha \hbar\right)\right]\psi^{*'}(x',t') \nonumber\\
&& =\psi^{*}(x',t').
\end{eqnarray}
Therefore the free particle propagator given by Eq. (\ref{ker-final}) is consistent with all the basic properties of a propagator.

\section{Conclusion}
In this paper, we have analyzed the deformation of the Heisenberg algebra consistent with both the GUP and DSR theory. 
It has been shown that this deformed Heisenberg algebra modifies the coordinate representation of the momentum operator,
which further modifies the Hamiltonian for all quantum mechanical systems. We have noted that this types of the more general
deformed Hamiltonian (\ref{ham-gen}) could contain fractional derivative terms. However, it was possible to give a formal 
meaning to these fractional derivative terms by using the theory of harmonic extension of functions. In fact, we also have 
constructed a formal expression for the quantum mechanical propagator corresponding to this Hamiltonian by using path 
integration. We have obtained the explicit form of quantum mechanical free particle propagator for one dimensional system. 
It has been observed that the evaluated propagator satisfies all the basic properties of a quantum mechanical propagator. 
Constructing plots of different probability 
amplitudes, we have shown that even in free particle case ordinary GUP and DSR theory shows different results. It will be possible to use  this propagator to  derive other quantities like the free particle partition functions and study its consequences.

It may be noted that recently this deformation of Heisenberg algebra has been used for analyzing various systems, such as,
transition rate of ultra
cold neutrons in gravitational field \cite{n6} and the Lamb shift and Landau levels \cite{n7}. In fact, it has been argued 
using this
algebra that the space is a discrete structure \cite{m1}. It would be interesting to analyze these systems by using path integral 
formalism
of quantum mechanics. It may also be observed, that in absence of matter fields, the Wheeler-DeWitt equation looks like the 
Schr\"{o}dinger
wave equation for a one dimensional particle with time dependent frequency \cite{wd,dw}. The modification of the 
Wheeler-DeWitt equation by
deforming the Heisenberg algebra has already been done \cite{mf}. The big bang singularity gets avoided in deformed 
Wheeler-DeWitt equation.
It would be interesting to
calculate the wave function of the universe corresponding to this deformed Wheeler-DeWitt equation using path integral
formalism.
An advantage of using path integral formalism for analyzing the deformed Heisenberg algebra is that this formalism can be 
generalized to field
theory, and then used for constructing
a deformed partition function in Euclidean quantum gravity. This in turn could be used for studding the deformation of 
spacetime foam.
It has been known that
it can be argued the cosmological constant vanishes due to the effects coming from spacetime foam  \cite{a10000,colm,colm2,b}.
However, we know from data obtained from type I supernovae that
our universe is accelerating in its expansion, and this mean that our universe has a small but finite  cosmological constant
\cite{supera,super1111,super2222,super3,super4,super5}. It would be interesting to repeat the analysis with a deformed
partition function for Euclidean quantum gravity. It might be possible to relate  this small but finite cosmological constant 
to the deformation
parameter of the Heisenberg algebra.
An alternative  deformation of the Heisenberg algebra generated by noncommutative spacetime has also been studied. 
In fact, this deformation has
been generalized to   non-anticommutativity, and this has in turn been applied to constructing models of non-local perturbative quantum gravity  \cite{nonanti,casdcasd , nonanti1}. It would   be interesting to study a combination non-anticommutativity with the deformation of the Heisenberg algebra produced by linear GUP. It would also been interesting to study the effects of this combined deformation on a free particle using path integral formalism. It may also be noted that deformation of field theories, consistent with GUP \cite{m, m1},
and a combination of GUP and  DSR \cite{dsr, dsr1},  have been studied. So, it would be interesting to analyze the combination of  these deformations of  field theories with  noncommutativity and non-anticommutativity.

\section*{Acknowledgments}
The research of Ahmed Farag Ali and Mohamed Moussa is supported by Benha University (www.bu.edu.eg).

\end{document}